\documentstyle[german,index,twoside,12pt]{article}
\begin{document}

\begin{titlepage}
\begin{flushright}
UWThPh-1997-32 \\
\today
\end{flushright}

\begin{center}

{\large \bf Ludwig Boltzmann -- A Pioneer of Modern Physics}
\footnote{Paper presented at the XXth International
Congress of History
of Science, on July 25, 1997, in Li\`ege, Belgium.}
 \\[0.5cm]
{\bf Dieter Flamm} \\[5mm]
Institut f\"ur Theoretische Physik der Universit\"at Wien, \\
Boltzmanngasse 5, 1090 Vienna, Austria \\
Email: Flamm@Pap.UniVie.AC.AT \\[1cm]
\end{center}
\renewcommand{\baselinestretch}{2.0}
\noindent
In two respects Ludwig Boltzmann was a pioneer of quantum mechanics.
First because in his statistical interpretation of the second law of
thermodynamics he introduced the theory of probability into a fundamental
law of physics and thus broke with the classical prejudice, that fundamental
laws have to be strictly deterministic. Even Max Planck had not been ready to
accept Boltzmann's statistical methods until 1900. With Boltzmann's pioneering
work the probabilistic
interpretation of quantum mechanics had already a precedent.
In fact in a paper in 1897
Boltzmann had already suggested to Planck to use his statistical methods for
the treatment of black body radiation. \\
The second pioneering step towards
quantum mechanics was Boltzmann's introduction of discrete energy levels.
Boltzmann used this method already in his 1872 paper on the $H$-theorem.
One may ask whether Boltzmann considered this procedure only as a 
mathematical device or whether he attributed physical significance to it.
In this connection Ostwald reports that when he and Planck tried to
convince Boltzmann of the superiority of purely thermodynamic methods
over atomism at the Halle Conference in 1891 Boltzmann suddenly said:
``I see no reason why energy shouldn't also be regarded as divided
atomically.'' \\
Finally I would like to mention, that Boltzmann in his lectures on Natural
Philosophy in 1903 already anticipated the equal treatment of space coordinates
and time introduced in the theory of special relativity. Furthermore
in the lectures by Boltzmann and his successor Fritz Hasen\"ohrl in Vienna the
students learned
already about noneuclidean geometry, so that they could immediately start to
work when Einstein's general theory of relativity had been formulated.
\end{titlepage}
Before 1900 classical Newtonian mechanics was the prototype of a
successful physical theory. As a consequence all physical laws had to be
strictly deterministic and universally valid. For most physicists --
among them Max Planck -- these rules applied also to thermodynamics.
They believed that the second law of thermodynamics was a basic axiom
handed down from God, which one had to accept as the starting point
of any thermodynamic consideration. In contrast to this view was Boltzmann's
statistical interpretation of the second law of thermodynamics which
he first formulated in 1877, nearly 50 years before the statistical
interpretation of quantum mechanics. Ludwig Boltzmann was born in
Vienna in 1844 and died in Duino near Trieste in 1906. He studied
mathematics and physics at the University of Vienna from 1863 to
1866, where Josef Stefan was his main advisor. He discussed a lot
with his senior friend Josef Loschmidt. \\

Boltzmann for the first
time introduced the theory of probability into a fundamental law of
physics and thus broke
with the classical prejudice, that fundamental laws of physics have to be
strictly deterministic. Because of this prejudice it is no wonder
that in the last decades
of the 19th century Boltzmann's statistical interpretation of thermodynamics
was not accepted by the physics community but met with violent
objections coming from physicists and mathematicians. These
objections were formulated in the form of paradoxes. The most
prominent objections were the reversibility paradox by Boltzmann's
friend Josef Loschmidt\footnote{J. Loschmidt: ``Uber den Zustand des
W\"armegleichgewichtes eines Systems von K\"orpern mit R\"ucksicht auf die
Schwerkraft, 1. Teil, Sitzungsber. Kais. Akad. Wiss. Wien Math.
Naturwiss. Classe {\bf 73} (1876) 128--142.} in 1876 and the recurrence paradox by 
E. Zermelo\footnote{E. Zermelo: ``Uber einen Satz der Dynamik und die
mechanische W\"armetheorie, Ann. Physik {\bf 57} (1896) 485--494.} in 1896.
Boltzmann met these objections by statistical arguments.
This struggle for Boltzmann's ideas, however, helped to pave the way
for the statistical interpretation of quantum mechanics in 1926 by
Max Born.
Boltzmann's expression for the entropy formulated in his
paper of 1877 entitled ''On the relation between the second law of
the mechanical theory of heat and the probability calculus with
respect to the theorems on thermal equilibrium''\footnote{ L.
Boltzmann: \"Uber die Beziehung zwischen dem Zweiten
Hauptsatze der mecha\-ni\-schen W\"armetheorie und der
Wahrscheinlichkeitsrechnung
resp. den S\"atzen \"uber das W\"armegleichgewicht, Sitzungsber.
Kais. Akad. Wiss. Wien Math. Naturwiss. Classe {\bf 76} (1877)
373--435.}, is a probabilistic
expression. He found that for a closed system the entropy $S$ of the
system is proportional to the phase space volume $\Omega$ occupied by the
macrostate of the system $ S \propto \log \Omega $. Furthermore
Boltzmann already introduced finite cells in phase space. Using such
cells to count the number of microstates this formula is now usually 
written in the notation of Max Planck $S = k \log W$. It implies
that the entropy $S$ is proportional to the logarithm of the
so called thermodynamic probability $W$ of the macrostate. $W$ is actually
the number of microstates by which the macrostate of the system can be
realized.
A macrostate is determined by a rather small number of macroscopic
variables of the system such
as volume, pressure and temperature. The latter two correspond to
averages over microscopic variables of the system. A microstate, on
the other hand, is specified by the
coordinates and momenta of all molecules of the system. Due to the large
number of molecules there is a very large number of different choices
for the individual coordinates and momenta which lead to the same macrostate.
For a closed system every microstate has the same a priori probability.
If one divides $W$ by the total number of microstates
accessible to the system including those of all other possible
macrostates one obtains the normalized probability to find a closed system
just in this macrostate. It turns out, that the largest number of
microstates corresponds to the state of thermodynamic equilibrium which
thus is the state of maximal entropy just as required by the second
law of thermodynamics. \\

Futhermore Boltzmann could show that for a very large system, by far
the largest number of all microstates corresponds to equilibrium-- and
quasi--equilibrium--states. The latter are states which differ
very little from the equilibrium state with maximum entropy and
cannot be distinguished macroscopically from the equilibrium state.
As an example let us create a nonequilibrium state by pouring, for
instance, a dye into a liquid. At first the dye will only be in a
limited region, which corresponds to a non\-equilibrium state, but it
will gradually spread through the whole liquid.
This behaviour is typical whenever you make such an experiment. It is
easy to tell the sequence in which snapshots of a
spreading dye were taken, even after their original order has been
deranged. How does this unidirectional behaviour in time come about?
In our example of a liquid containing a dye the equilibrium-- and
quasi--equilibrium--states correspond to a
practically uniform distribution of the dye through the whole liquid
with very small intensity fluctuations of the dye. With increasing size
of the system, the preponderance of the equilibrium- and
quasi-equilibrium-states becomes even more overwhelming. If for every
possible microstate of a large system we put a marked sphere into an urn
and afterwards drew spheres from the urn indiscriminately, we would
practically always draw an equilibrium-- or quasi--equilibrium--state.
The transition from nonequilibrium to equilibrium thus corresponds to
a transition from exceptionally unprobable nonequilibrium-states to
the extremely probable equilibrium--state. This is Boltzmann's statistical
interpretation of the second law. The appearance of so--called statistical
fluctuations in small subsystems was predicted by Boltzmann and he
recognized Brownian motion as such a phenomenon. The theory of
Brownian motion has been worked out independently by Albert Einstein
in 1905 and by Marian Smoluchowski. The experimental verification of
these theoretical results by Jean Baptiste Perrin was important
evidence for the existence of molecules. \\
   
The second pioneering step which Boltzmann made towards quantum
mechanics was the introduction of discrete energy levels for
molecules contained in a finite volume. Boltzmann used this method
already in his paper of 1872 on the $H$--theorem\footnote{L.
Boltzmann: Weitere Studien \"{u}ber das W\"{a}rmegleichgewicht
unter Gasmolek\"{u}len, Sitzungsber. Kais. Akad. Wiss. Wien Math.
Naturwiss. Classe {\bf 66} (1872) 275--370.} and he needed it in
the above mentioned paper of 1877 to be able to enumerate the number
of microstates. Boltzmann actually introduced cells of finite size in
phase--space spanned by the coordinates and momenta of all molecules.
In quantum mechanics Heisenberg's uncertainty principle fixes the
minimal size of such cells. Max Planck was certainly influenced by
these ideas of Boltzmann. \\

In 1897 Boltzmann had a dispute with Planck on the irreversibility
of radiation phenomena which may have stimulated Planck's discovery of
quantum mechanics in 1900. At this time Planck thought that he could
derive irreversible
behaviour for radiative processes without any assumptions on the initial
states. Boltzmann could, however, show that this was not true and at the
beginning of his second paper answering Planck\footnote{L. Boltzmann:
\"Uber irreversible Strahlungsvorg\"ange II., Berliner Ber. (1897)
1016--1018.} he made the following suggestion to Planck: \\

``It is certainly possible and would be gratifying to derive for
radiation phenomena a theorem analogously to the entropy theorem from the
general laws for these phenomena using the same principles as in gas
theory. Thus I would be pleased, if the work of Dr. Planck on the
scattering of electrical plane waves by very small resonators would become
useful in this respect, which by the way are very simple calculations whose
correctness I have never put in doubt.

Only if Dr. Planck in his second communication claims again that no other
process in nature is known, in which conservative forces lead to irreversible
changes, I can not agree.'' \\

Indeed Planck followed Boltzmann's recommendation and used Boltzmann's
statistical methods for the derivation of his celebrated law for the
black body radiation. Planck used thereby the additional assumption that
classical oscillators absorb and emit energy only in integer multiples of
the product of Planck's constant $h$ with the frequency $\nu$ of the
radiation which gave rise to the birth of quantum mechanics. In fact,
in the framework of Boltzmann's statistical approach it was quite common
to introduce discrete energy levels to obtain a denumerable set of states.
Boltzmann used this method already in his 1872 paper on the
$H$-theorem\footnote{L. Boltzmann: Weitere Studien \"uber das
W\"armegleichgewicht unter Gasmolek\"ulen, Sitzungsber. Kais. Akad. Wiss.
Wien Math. Naturwiss. Classe {\bf 66} (1872) 275--370.}.
One may ask whether Boltzmann considered this procedure only as a 
mathematical device or whether he attributed physical significance to it.
In this connection Ostwald reports that when he and Planck tried to
convince Boltzmann of the superiority of purely thermodynamic methods
over atomism at the Halle Conference in 1891 Boltzmann suddenly said:
\\

``I see no reason why energy shouldn't also be regarded as divided
atomically.''\footnote{W. Ostwald: Lebenslinien -- Eine
Selbstbiographie, Klasing, Berlin 1927, vol. 2 p. 187 and 188.} \\

At this time Planck was still an opponent of the atomistic theory. In
1900, however,
he was converted from ``Saulus'' to ``Paulus'' when he had to use 
Boltzmann's statistical methods to explain his law of radiation  and
he became the most important proponent of Boltzmann's ideas. After
Boltzmann's death in 1906 it was the authority of Max Planck and
Albert Einstein which brought general recognition of Boltzmann's
ideas. In the years 1909 and 1910, for instance, Planck had a vivid
dispute with Ernst Mach on the existence of atoms. Without the concept
of atoms it is very unlikely that quantum theory would have
developed. The existence of atoms is one of the basic assumptions of
Boltzmann's statistical mechanics for which he fought vigorously.
In fact the deviations from the classical equipartition theorem for
the energy observed for black body radiation and for the specific
heat of solids were important evidence for quantum theory. Especially
Erwin Schr"dinger, the founder of wave mechanics, studied the
applications of Boltzmann's statistical methods in great detail. A
manifestation of Boltzmann's influence on Schr\"odinger is
Schr\"odinger's enthusiastic quotation of Boltzmann's line of thought: \\

``His line of thought may be called my first love in science. No
other has ever thus enruptured me or will ever do so
again.''\footnote{E. Schr\"odinger, S.B. Preuss. Akad. Wiss. Berlin
(1929), pp. C--CII; reprinted in E. Schr"dinger, Science Theory and
Man, (New York: Dover 1957), p. XII.} \\

Finally I would like to mention that Boltzmann in his lectures on
Natural Philosophy in 1903 at the University of Vienna as well as in
the second volume of his textbook on classical mechanics already
anticipated the equal treatment of space coordinates and time which is
introduced in the special theory of relativity. Furthermore in the
lectures of Boltzmann and of his successor Fritz Hasen"hrl in Vienna
the students learned already about noneuclidean geometry, so that
they could immediately start to work in this field when Einstein's
general theory of relativity had been formulated. A result of this
work was, for instance, the paper by Hans Thirring and Josef
Lense\footnote{H. Thirring and J. Lense: \"Uber den Einflu\3 der
Eigenrotation der Zentralk\"orperauf die Bewegung der Planeten und
Monde nach der Einsteinschen Gravitationstheorie, Phys. Zeitschrift,
Leipzig Jg.19 (1918), No. 8, p. 204--205.} on the effects of the
rotation of the earth on the moon and on satellites.

\end{document}